# Phase boundaries of nano-dots and nano-ripples over a range of collision cascades

Oluwole Emmanuel Oyewande

Department of Physics, University of Ibadan, Ibadan

***ABSTRACT***

*One of the open problems of the formation and evolution of the nanostructures on solid surfaces that are driven by particle irradiation is whether all the nano-patterns can be accounted for by the continuum theory, since the Cuerno-Barabasi continuum theory explanations have thus far focused mainly on ripple patterns and rough self-affine scaling. In this article, we extend continuum theoretical calculations based on the nonlinear Cuerno-Barabasi theory to regions yet unexplored in the continuum theory literature but recently shown, in Monte Carlo simulations, to be devoid of ripple patterns. We obtained results of weighted ion penetrations $a_\alpha$ and $a_\rho$ for different impingement angles. Our results show that the balance of sputtering coefficients required for ripple patterns was never attained in this region, which confirmed that ripple patterns are indeed absent and that the continuum theory does not fail as previous studies indicated. In particular our results show a remarkable agreement between the continuum and the discrete (solid-on-solid) models; by calculations based on the continuum theory cascade parameters alone, we obtained the same topographic phase diagram predicted in the recent Monte Carlo simulation studies.*

**Keywords**: *Topographic phase diagrams; nanostructures; nano-dots; Cuerno-Barabasi model.*

## 1.0 INTRODUCTION

The irradiation of a solid surface at sub-micrometer length scales is known to lead to the formation of ordered patterns on the surface. This phenomenon has attracted a great deal of interest and scientific activities in the material science community for quite some time and more so in recent years [1; 2; 3; 4; 5; 6]. The interest is partly due to the tuneable characteristics of the nanostructures which makes them applicable in nano-device and opto-electronic applications, and partly due to the influence an understanding of them might wield in related fields of investigation [1, 7].

Much is now understood about the formation and dynamics of the ordered nanostructures. For instance the linear theory of Bradley and Harper [2] explains the formation of ripple topographies as a result of the curvature dependence of the sputter yield which leads to an instability of the surface against the sputtering process, that is, the ejection of surface particles by the irradiated particles, such that surface depressions are eroded in preference to surface protrusions. The linear regime, where the Bradley-Harper theory is relevant is believed to be at the initial stages of the sputtering process. As time goes on, nonlinear effects arising from increasing local surface gradients are no longer negligible and the linear theory breaks down. Cuerno and Barabasi [5] proposed a noisy nonlinear partial differential equation to account for the nonlinear effects. Unlike the deterministic Bradley-Harper model, the Cuerno-Barabasi model is stochastic and its noise term accounts for the random nature of the sputtering events. The nonlinear model was introduced to address ripple formation and absence of ripple formation in sputtering conditions where the linear theory breaks down, and any topography beyond these were considered unusual and beyond the theory; which has lead to the ever-increasing inclusion of additional terms to the original partial differential equation of Cuerno and Barabasi.

More recent advances in surface probe techniques and investigations have revealed a richer variety of surface topographies [7]. Since dynamical systems can always be modelled by partial differential equations whose solutions can be of infinite variety and strongly subject to and dependent on the initial conditions, deviations from the predictions of the nonlinear theory may be due to unexpected relative initial conditions arising in the course of the sputtering, for instance, with the noise as a possible influence. Hence, in this article we applied the Cuerno-Barabasi model in its present form without additional terms, since this form already contains the entire physics of the problem. We explore a region of current interest brought into focus by Monte Carlo simulations [8; 9] which reveal the formation of off-normal incidence nano-dots instead of nano-ripples, in order to address the open problem of whether the continuum theory fails in this region, and if it does fail, whether additional terms are necessary.

We provide phase diagram calculations of the continuum theory over a range of collision cascade parameters corresponding to an anisotropic distribution of the energy of the impinging ions. In particular, we provide the results of the continuum theory as regards nano-dots. The rest of this paper is organized as follows. In the next section we discuss the theory. In section 3.0, we present and discuss our results for the anisotropic cases relevant to the regions studied in the Monte Carlo simulations. Finally, we provide a brief conclusion in section 4.0.

## 2.0 MATERIALS AND METHOD

When a solid surface is irradiated by a beam of ions, the surface energy is increased and atoms in the vicinity of the ion impact may gain sufficient energy to overcome the binding energy barriers that confine them to the material. According to P. Sigmund, the energy of the impinging ion follows a Gaussian distribution which peaks at the point of impact and tapers to zero away for the impact site. Thus, the distribution $E(x)$ of the energy $E$ of the incident ion to a surface particle located at position $x = (x_1, x_2, x_3)$ is assumed to be of the Gaussian form [10]:

$$E(x) = \frac{E}{\left(\sqrt{2\pi}\right)^3 \alpha \rho^2} \exp\left[-\frac{\rho^2 x_3^2 + \alpha^2(x_1^2 + x_2^2)}{2\alpha^2 \rho^2}\right], \qquad (1)$$

where $\alpha$ and $\rho$ are the widths of the distribution parallel and perpendicular to the ion beam direction, respectively. The erosion velocity $v \propto \partial_t h$ by definition, following which the dynamic evolution of the surface height $h(x,t)$ at nanometre length scales is for most cases governed by the Cuerno-Barabasi nonlinear continuum model [2; 5]

$$\partial_t h(x,t) = -v_0 + \zeta \partial_{x_1} h(x,t) + \varsigma_{x_1} \partial_{x_1 x_1} h(x,t) + \varsigma_{x_2} \partial_{x_2 x_2} h(x,t)$$

$$+\eta_{x_1}[\partial_{x_1} h(\boldsymbol{x},t)]^2 + \eta_{x_2}[\partial_{x_2} h(\boldsymbol{x},t)]^2 - D\nabla^4 h(\boldsymbol{x},t) + \beta. \quad (2)$$

$v_0$ is the erosion velocity of a flat surface, $\zeta$ is a proportionality constant related to the local surface slope along the x-direction, $\varsigma_{x_1}$ and $\varsigma_{x_2}$ are the (linear) surface tension coefficients, $\eta_{x_1}$ and $\eta_{x_2}$ are the nonlinear coefficients, $D$ is the surface diffusion coefficient, $\beta$ is a (Gaussian) noise term with zero mean, representing the randomness in the ejection of the surface constituents.

Using the convenient notation,

$$a_\alpha = \frac{a}{\alpha}, a_\rho = \frac{a}{\rho}, \kappa = \cos\theta, \sigma = \sin\theta, \varpi = a_\alpha^2 \sigma^2 + a_\rho^2 \kappa^2, \Upsilon = \frac{FPa}{a\rho\sqrt{2\pi\varpi}} \exp\left(-\frac{a_\alpha^2 a_\rho^2 \kappa^2}{2\varpi}\right),$$

where $F$ is the ion flux and $P$ is the proportionality constant between the power deposition and the rate of erosion (we set $FEP = 1$), we provide the (Cuerno-Barabasi) coefficients, for ease of reference, as follows [5; 6]:

$$\varsigma_{x_1} = \Upsilon a \frac{a_\alpha^2}{2\varpi^3}(2a_\alpha^4 \sigma^4 - a_\alpha^4 a_\rho^2 \sigma^2 \kappa^2 + a_\alpha^2 a_\rho^2 \sigma^2 \kappa^2 - a_\rho^4 \kappa^4), \qquad \varsigma_{x_2} = -\Upsilon a \frac{\kappa^2 a_\alpha^2}{2\varpi}$$

$$\eta_{x_1} = \Upsilon \frac{\kappa}{2\varpi^4}\big[a_\alpha^8 a_\rho^2 \sigma^4(3 + 2\kappa^2) + 4a_\alpha^6 a_\rho^4 \sigma^2 \kappa^4 - a_\alpha^4 a_\rho^6 \kappa^4(1 + 2\sigma^2)\big]$$
$$- \varpi^2\big[2a_\alpha^4 \sigma^2 - a_\alpha^2 a_\rho^2(1 + 2\sigma^2)\big] - a_\alpha^8 a_\rho^2 \sigma^2 \kappa^2 - \varpi^4 \qquad (3)$$

$$\eta_{x_2} = \Upsilon \frac{\kappa}{2\varpi^2}(a_\alpha^4 \sigma^2 + a_\alpha^2 a_\rho^2 \kappa^2 - a_\alpha^4 a_\rho^2 \kappa^2 - \varpi^2)$$

We obtained the phase diagrams of the continuum theory for both symmetric and asymmetric energy distribution cases in terms of the weighted variables $a_\alpha$ and $a_\rho$, each ranging from 0 to 5. Our results therefore include some collision cascade ranges of interest studied in recent reports [8; 9] of Monte Carlo simulations of surface sputtering in which atomistic simulations [11] of the range of ions in solids were performed to determine the realistic ranges of collision cascade parameters used in the MC simulations. Our results are presented and discussed in the next section.

**3.0 RESULTS AND DISCUSSION**

The values of the sputter coefficients $\varsigma_{x_1}, \varsigma_{x_2}, \eta_{x_1}, \eta_{x_2}$ as functions of the dimensionless quantities $a_\alpha$, and $a_\rho$ have been used to obtain the phase diagrams presented. The phase boundaries are the projections of the points of intersection of the coefficients, to within an infinitesimal positive amount $\delta$, onto the $(a_\alpha, a_\rho)$ plane; points for which, e. g. $|\varsigma_{x_1} - \varsigma_{x_2}| < \delta$. For ripple patterns to be formed $\varsigma_{x_1}$ and $\varsigma_{x_2}$ must be less than zero and the wavelength of the ripples is given by $\lambda = 2\pi\sqrt{2D/|\varsigma|}$ and their orientation is along the direction with the largest absolute value $|\varsigma|$.

Figure 1 shows the phase diagram for the different possible topographic phases defined by the dimensionless quantities $a_\alpha$ and $a_\rho$ ranging from 0 to 5. From left to right $\theta = 50°$, $\theta = 70°$, and $\theta = 30°$, respectively. More regions are dominant in the phase diagram as found in recent studies [Refs. 12 – 14], the prominent ones from top to bottom of Fig. 2 are defined (see Ref. 14), respectively, by the combinations: $\varsigma_{x_1} > 0, \varsigma_{x_2} < 0, \eta_{x_1} > 0, \eta_{x_2} < 0$ (dot);

$\varsigma_{x_2} < \varsigma_{x_1} \leq 0, \eta_{x_1} > 0, \eta_{x_2} < 0; \varsigma_{x_2} < \varsigma_{x_1} \leq 0; \eta_{x_1} < 0, \eta_{x_2} < 0$. Due to the influence of the nonlinearities the first region is not characterised by ripples, i.e. since $\varsigma_{x_1}$ is positive. This region is the one for which nano-dots have been found in the Monte Carlo simulations.

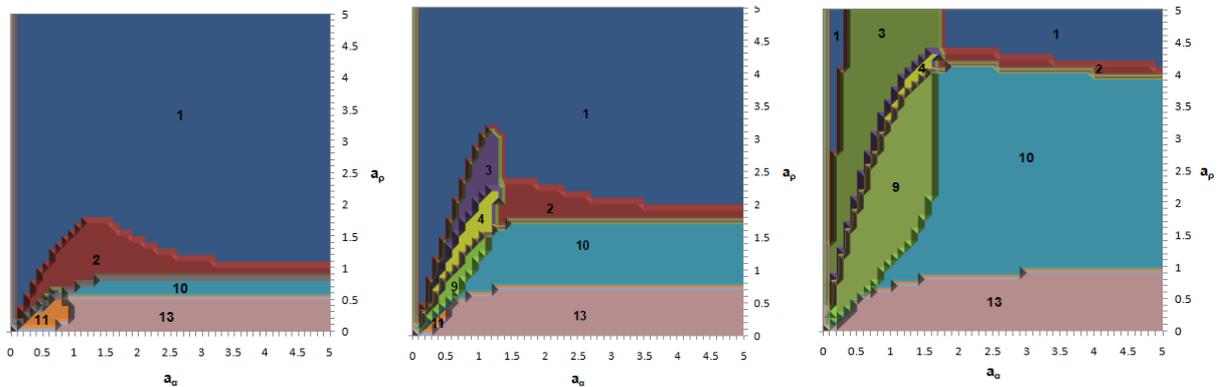

Figure 1: Phase diagram for the different possible topographic phases defined by the dimensionless quantities $a_\alpha$ and $a_\rho$ ranging from 0 to 5. L-R: $\theta = 30°$, $\theta = 50°$, $\theta = 70°$. The regions of the phase diagram are as defined in the text. The figures show the boundary shifts in the phase diagram, arising from varying $\theta$.

Thus, the continuum theory does not fail in this particular geometry of the scattering cascades of the impinging ion in the first few surface layers. In addition to this result, figure 1 also shows shifts in the phase boundaries as $\theta$ varies with the result that periodic ripple structures may occur even where nano-dots has been found provided that $\theta$ is sufficiently small.

The figures [Fig. 1(a) – (c)] show a striking agreement with recent Monte Carlo simulations of a discrete solid-on-solid model of sputter erosion [8]. Furthermore, the boundary shifts observed in the simulation are present here as well; Figure 1 shows that as $\theta$ increases the phase boundaries expand outwards. The rich variety of topographic regions, and the consequent boundaries, reduce if only the collision cascade parameters $\alpha$ and $\rho$ are used and not weighted by the penetration depth $a$; this is shown in Figure 2.

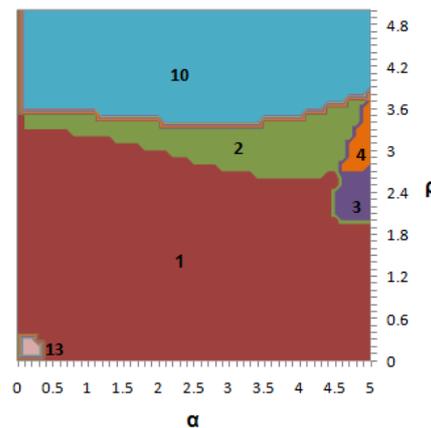

Figure 2: Phase diagram for the different possible topographic phases defined by the quantities $\alpha$ and $\rho$ ranging from 0 to 5.

Note that the orientation of the boundaries is reversed (top-bottom) in Figure 2 relative to Figure 1 because Fig. 1 relates to the inverse of these cascade parameters.

Both figures demonstrate that the balance of sputtering coefficients required for ripple patterns is never attained in the dot region, which confirms that ripple patterns are indeed absent and that the continuum theory does not fail as previous studies indicate.

**4.0 CONCLUSION**

Topographic phase diagram calculations over a range of collision cascade parameters of impinging ions in a sputtered surface have been presented. The range, being representative of different distributions of the energy of the impinging ion in the surface, encompasses a variety of sputtering scenarios for different materials. The results reveal the occurrence of non-ripple nanostructures in the Cuerno-Barabasi continuum theory, contrary to the previously held belief that the Cuerno-Barabasi model only predicts ripple topographies and self-affine rough surfaces. Furthermore, our continuum theory calculations reveal very good agreement with the Monte Carlo simulation results. Moreover, we found that nano-dots predicted by both theories for off-normal incidence sputtering may not be observed if the incidence angle is too small.

e-mail: eoyewande@gmail.com